\PassOptionsToPackage{unicode=true}{hyperref} 
\PassOptionsToPackage{hyphens}{url}
\documentclass[]{article}
\usepackage{lmodern}
\usepackage{amssymb,amsmath}
\usepackage{ifxetex,ifluatex}
\usepackage{fixltx2e} 
\ifnum 0\ifxetex 1\fi\ifluatex 1\fi=0 
  \usepackage[T1]{fontenc}
  \usepackage[utf8]{inputenc}
  \usepackage{textcomp} 
\else 
  \usepackage{unicode-math}
  \defaultfontfeatures{Ligatures=TeX,Scale=MatchLowercase}
\fi
\IfFileExists{upquote.sty}{\usepackage{upquote}}{}
\IfFileExists{microtype.sty}{%
\usepackage[]{microtype}
\UseMicrotypeSet[protrusion]{basicmath} 
}{}
\IfFileExists{parskip.sty}{%
\usepackage{parskip}
}{
\setlength{\parindent}{0pt}
\setlength{\parskip}{6pt plus 2pt minus 1pt}
}
\usepackage{hyperref}
\hypersetup{
            pdfborder={0 0 0},
            breaklinks=true}
\usepackage{longtable,booktabs}
\IfFileExists{footnote.sty}{\usepackage{footnote}\makesavenoteenv{longtable}}{}
\providecommand{\tightlist}{%
  \setlength{\itemsep}{0pt}\setlength{\parskip}{0pt}}
\ifx\paragraph\undefined\else
\let\oldparagraph\paragraph
\renewcommand{\paragraph}[1]{\oldparagraph{#1}\mbox{}}
\fi
\ifx\subparagraph\undefined\else
\let\oldsubparagraph\subparagraph
\renewcommand{\subparagraph}[1]{\oldsubparagraph{#1}\mbox{}}
\fi

\usepackage{graphicx}
\usepackage{pdfpages}

\makeatletter
\def\fps@figure{htbp}
\makeatother

\date{}

 \usepackage{fancyhdr}

\begin{document}

\textbf{Disentangling the socio-ecological drivers behind illegal
	fishing in a small-scale fishery managed by a TURF system} \footnote{© 2020. This manuscript version is made available under the CC-BY-NC-ND 4.0 license \href{http://creativecommons.org/licenses/by-nc-nd/4.0/}{http://creativecommons.org/licenses/by-nc-nd/4.0/}}

\textbf{}\\

Silvia de Juan\textsuperscript{1*}, Maria Dulce
Subida\textsuperscript{2}, Andres Ospina-Alvarez\textsuperscript{3},
Ainara Aguilar\textsuperscript{2}, Miriam Fernandez\textsuperscript{2}

\textsuperscript{1} Spanish Scientific Research Council, Institute of
Marine Sciences (ICM-CSIC), Passeig Maritim de la Barceloneta 37-49,
Barcelona, Spain.

\textsuperscript{2} Estacion Costera de Investigaciones Marinas,
Pontificia Universidad Catoolica de Chile, Alameda 340, C.P. 6513677,
Casilla 193, Correo 22, Santiago, Chile.{~}

\textsuperscript{3} Spanish Scientific Research Council, Mediterranean
Institute for Advanced Studies (IMEDEA-CSIC/UIB), C/ Miquel Marques 21,
CP 07190 Esporles, Balearic Islands, Spain.

*corresponding author: sdejuan@icm.csic.es

\textbf{Highlights:{~}}

\begin{itemize}
\tightlist
\item
  This study describes the development of a Bayesian network adapted to
  complex SSF.
\item
  A Bayesian network was designed to identify drivers of illegal
  fishing.
\item
  The network was based on the link between socio-economic drivers and
  resource state.
\item
  Scenario analysis explored effects of variables that are susceptible
  to be managed.{~}
\end{itemize}

\textbf{Abstract:} A substantial increase in illegal extraction of the
benthic resources in central Chile is likely driven by an interplay of
numerous socio-economic local factors that threatens the success of the
fisheries' management areas (MA) system. To assess this problem, the
exploitation state of a commercially important benthic resource (i.e.,
keyhole limpet) in the MAs was related with socio-economic drivers of
the small-scale fisheries. The potential drivers of illegal extraction
included rebound effect of fishing effort displacement by MAs, level of
enforcement, distance to surveillance authorities, wave exposure and
land-based access to the MA, and alternative economic activities in the
fishing village. The exploitation state of limpets was assessed by the
proportion of the catch that is below the minimum legal size, with high
proportions indicating a poor state, and by the relative median size of
limpets fished within the MAs in comparison with neighbouring OA areas,
with larger relative sizes in the MA indicating a good state. A
Bayesian-Belief Network approach was adopted to assess the effects of
potential drivers of illegal fishing on the status of the benthic
resource in the MAs. Results evidenced the absence of a direct link
between the level of enforcement and the status of the resource, with
other socio-economic (e.g., alternative economic activities in the
village) and context variables (e.g., fishing effort or distance to
surveillance authorities) playing important roles. Scenario analysis
explored variables that are susceptible to be managed, evidencing that
BBN is a powerful approach to explore the role of multiple external
drivers, and their impact on marine resources, in complex small-scale
fisheries.{~}


INTRODUCTION

Small-scale, artisanal, fisheries (SSF) are {the pillar of wellbeing for
many coastal communities, as it is estimated that they contribute to
half of the global catch {[}1{]}, while employing 90\% of the world's
fisheries {[}2{]}. However, these fisheries are largely unassessed
{[}3{]}} and often data poor {[}4{]}. SSF are becoming a priority for
FAO, and efforts are on improving data gathering to guide monitoring and
management protocols. Innovative approaches to maximize the available
information from SSF are critical, e.g., the incorporation of
Traditional Ecological Knowledge {[}5{]} or the use of fishers'
perception surveys {[}6{]}. While data gathering for the assessment of
SSF is improving, there are several caveats driving fish stocks to
overexploitation that need urgent assessment. These factors may be
classified into two interacting types: those related with failure of
management and conservation rules {[}7{]} and those related with the
socio-economic context as the loss of the culture of care and
responsibility for marine resources prompted by strongly market oriented
fishery policies {[}8{]}.{~}

Scientific evidence suggests that non-compliance with fishing
regulations is a widespread phenomenon in global fisheries {[}3,9,10{]}
, and emerges among the most important factors contributing to the
overexploitation of marine resources {[}11,12{]}. However, information
on illegal fishing practices in SSF is still scarce. SSF are potentially
more prone to violations to catch limits or minimum sizes due to the
nature of the operation {[}13{]}. Their usual spatially scattered nature
impose serious challenges to monitoring, surveillance and enforcement to
detect any non-compliance activities, facilitating illegal fishing
{[}3,9,10{]}. It is a complex problem to solve, as numerous
socio-economic factors are probably playing a key role, and it is likely
to be highly conditioned by local context characteristics {[}8,14{]}.{~}

The absence of incentives to comply with fisheries normative has been
pointed out as a significant problem for SSF, particularly in scenarios
where neoliberal fishing policies fostered severe extractivism in
detriment of traditional and more sustainable fisheries ({[}8{]} and
references therein). The common pool resources dilemma {[}15{]}{~ }has
been in part solved using Territorial User Right in Fisheries (TURF).
While these co-management systems aim to encourage trust, rule
compliance and fishers' involvement in enforcement {[}16{]}, they are
also likely to allow fishers to sell fish at higher prices, reduce
resource waste and increase fishers' incomes ({[}17{]} and references
therein). Chile was pioneer on the implementation of a TURF system to
the SSF at a national scale in 2003, known as Management and
Exploitation Areas for Benthic Resources (AMERB, hereafter MA) {[}18{]}.
The TURF system in Chile has contributed in increasing abundance of
commercial species {[}19{]} and has shown positive effects on
biodiversity and trophic web structure {[}20,21{]}. However, the fishing
pressure outside these management areas {[}22{]} and a negative
perception of fishers on the TURF system in aspects related with
economical revenues and government support {[}23,24{]} urge an upgrade
in this management paradigm. Recent {studies suggest that illegal
extraction in the TURFs can be as high as 68\% of the annual income
obtained from this system in some regions of Chile {[}25{]}, and recent
biological surveys provided evidences of poaching in several management
areas {[}26{]}. Research on fisheries working on a TURF basis, including
some studies }in central Chile,{ identified }a series of incentives
promoting illegal fishing. For instance, the usually limited human
resources available for surveillance activities (both from the
authorities and fishers) make it difficult to identify and punish
poachers, and that there is a well-established black market in demand of
fisheries products {[}27,28{]}. {These findings trigger the need of an
integrative approach to identify which are the main drivers of illegal
extraction in Chile TURF system.{~}}

Usually, TURFs' efficiency is assessed through the ecological state of
biological components or through the socio-economic benefits of the
management {[}17{]}; the interaction between these two aspects has
seldom been addressed. Here, we provide a quantitative assessment of the
effects of socio-ecological drivers on the illegal fishing of the
traditional and culturally important keyhole limpet fishery in the
central coast of Chile. A Bayesian Belief Network (BBN) model was
developed linking key biological and socio-economic components of the
SSF. BBNs have been applied to consider management systems governing
artisanal fisheries where the effects of qualitative and quantitative
factors are of concern {[}29--33{]}, or when considering social,
environmental and economic factors leading to multi-objective management
of coastal resources {[}34,35{]}. In the present work, we used a BBN
model that integrates data from fisheries stakeholders and scientists to
identify the key drivers that influence the proportion of illegal
fishing in the TURF system, as well as the relationship between factors
that have generally been regarded as determining the state of the
resource (e.g., level of enforcement) and contextual factors (e.g.,
rebound effect of fishing effort displacement or distance to
surveillance authorities). The BBN approach can help building a
socio-ecological model of fisheries in an environment that is partly
data-poor. Moreover, the graphical outputs also facilitate communicating
the results to stakeholders. The present paper addresses relevant issues
driving illegal extraction in the TURF system, including the
identification of a) factors that determine the level of effective
enforcement needed to reduce illegal fishing, and b) variables that can
assist rapid assessment of effective enforcement and success of the SSF
system.

 METHODS

\textit{Small-scale fisheries in central Chile}

Small-scale benthic fisheries in Chile are mostly organized around
fishing coves locally known as \emph{caletas}, which serve as
operational bases for the local fleet and fisher organizations. The
benthic fisheries operate around 17 km away from the coves {[}22{]} to a
depth of approximately 20 m (official diving depth), and the harvesting
fishing grounds show two contrasting management regimes: (i) exclusive
harvest rights assigned to fishers' organizations (TURFs), locally known
as management areas (hereafter MAs) or (ii) historical fishing grounds
without spatial entry restrictions, hereafter referred as open access
areas (OAs). {The most common fishery resources that can be extracted in
the MAs are the Chilean abalone or loco (\emph{C. conholepas}), keyhole
limpets (a set of species of the genus \emph{Fissurella}), the red sea
urchin (\emph{Loxechinus albus}) and kelp (mainly \emph{Lessonia} spp.)
that are exclusively exploited by fishermen pertaining to the fisheries
association. Outside the MAs, both fishers belonging to the association
and officially registered un-associated fishers can extract fish and
benthic resources (except loco). }While some species-specific
regulations operate for both management regimes (e.g. temporal
reproductive bans or minimum legal size) others apply exclusively to MAs
(annual quotas) or OA (total ban of locos). {However, due to differences
in the administration of the MAs, not all these areas exhibit a similar
enforcement level {[}36{]}. A well-enforced MA implies that the maximum
quotas are respected, and the area is surveyed }to avoid illegal
fishing. The control of catches and enforcement of regulations is
usually performed by the National Fisheries Service, with about 25,000
enforcement actions during 2018 focusing on MLS regulations in both
artisanal and industrial fleet {[}24{]}. However, {fishers' associations
must cover surveillance costs, although it is not mandatory. In the
present work, the term poaching refers to the illegal extraction of the
resource, either due to noncompliance with the annual quotas by the
associated fishermen, or to illegally extracting the resource from MAs
by un-associated fishermen.}

For the present study, data was gathered from 13 fishers' associations
that manage 24 areas in the central coast of Chile (Fig. 1). Each
fishers' association presented between 1 and 3 operational MAs at the
time of the study. Different sources of data were gathered for the
study. In 2016-2020, face-to-face interviews and telephonic
questionnaires were carried out to the leaders of the fishers'
associations to obtain information and perceptions on MA enforcement,
poaching intensity and the organization of the fishers association. In
2017-2018, biological surveys were conducted in MAs and neighbouring OAs
to assess keyhole limpet size structure.{~}

\begin{figure}
	\centering
	\includegraphics[width=0.9\linewidth]{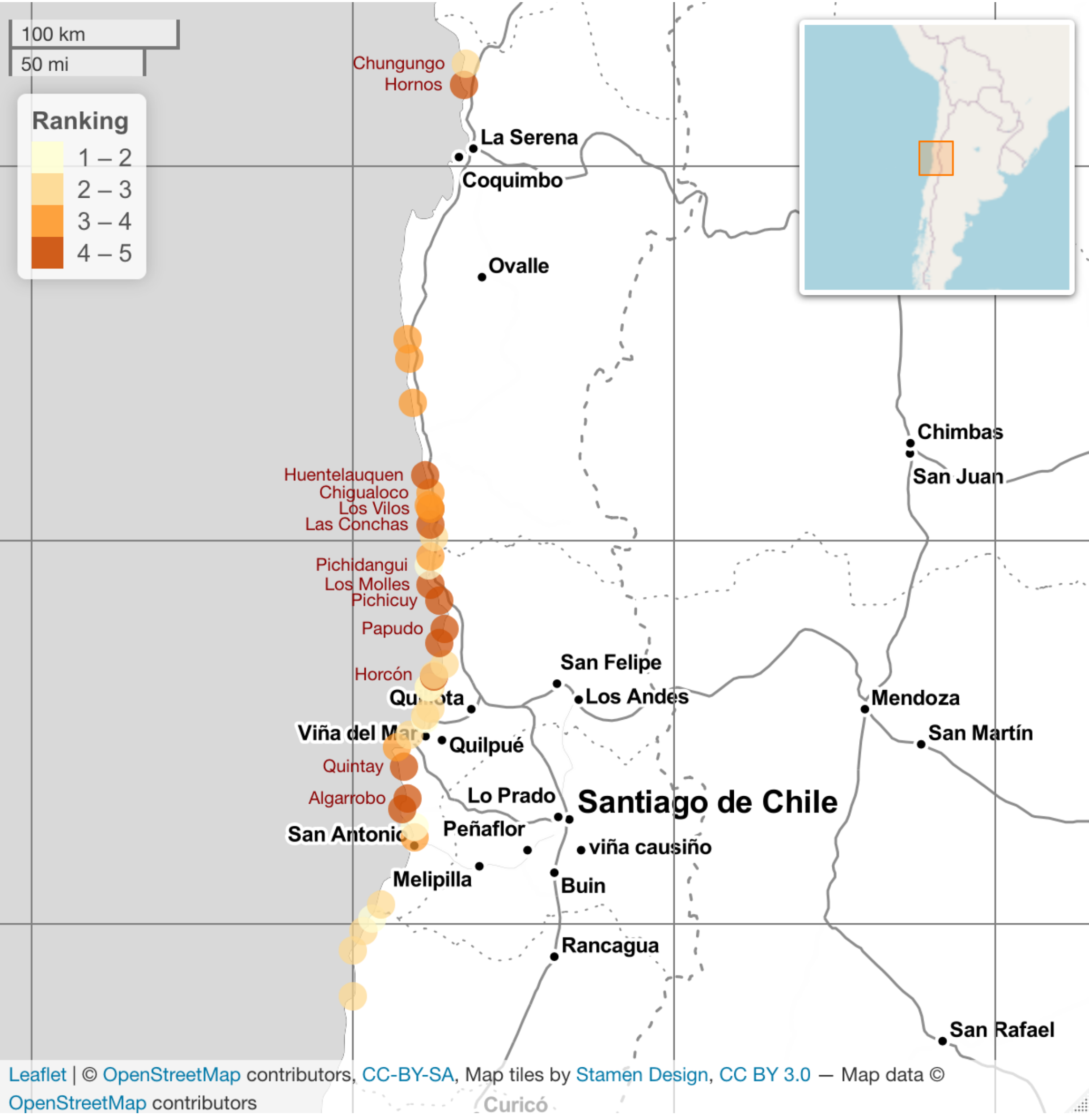}
	\caption{Map of the study area including the names of the 13 coves included in the study (local names in red). Note that each association might have between 1 and 2 management areas, resulting in 24 areas included in the analysis. The ranking corresponds to the enforcement level (5 being the highest enforcement). The ranking was performed with a wider set of 39 fishers' associations (included in the map).}
	\label{fig:fig-1}
\end{figure}

\textit{Interviews with fishers}

Interviews with the leaders of the fishers' associations were carried
out from 2016 to 2020 in 40 associations along the central coast of
Chile. Two sets of interviews were carried out: (a) face-to-face
extended interviews focused on 16 coves that covered ca. 250 km of the
coast in central Chile; and (b) remote (telephonic) interviews focused
on 24 coves covering the entire study area (\emph{ca.} 700 km). The two
approaches used the same questionnaire that aimed to explore the
enforcement protocol, or lack of it, the evenness in enforcement across
MAs and the fishers' perception on the enforcement effectiveness and
poaching records per year. The full questionnaire is provided in
supplementary material.{~}

The information gathered during these interviews allowed defining the
level of enforcement endured by each fishers' association. The
enforcement ranking was based on a dichotomy tree that represented all
the options that a fishers' association could exhibit. As a result, an
area could have a very high level of enforcement (rank 5), when the
fisher association hires an external person to perform 24h surveillance;
on the other end, a very low enforcement (rank 1) occurs when there is
no surveillance in the area (Fig. 2). Differences between high and very
high enforcement, with 24 hours surveillance performed by either the
members of the fishermen association or by a payed person respectively,
is justified by an expected higher compromise by the paid person
{[}23{]}. Fishers perception on enforcement effectiveness (i.e.,
fishermen perceived the enforcement in their MAs was either effective or
non-effective) and surveillance differences among the MAs controlled by
each fishers' association (i.e., an association survey all MA equally
\emph{vs}. surveillance efforts focus on a single MA) allowed the
inclusion of an additional variable that considered the
``effectiveness'' of enforcement. To compute this variable, the
enforcement level of a MA (ranked in 1 to 5) was reduced by one score if
the fishers' association dedicated less time to survey this MA (relative
to other MAs controlled by the same association), and by an additional
score if the fishers perceived that enforcement in their MAs was not
effective. As a result, 20 out of the 24 MAs had an ``effective
enforcement'' lower than the formal ``enforcement level''. During the
interviews, fishers also provided information on number of poaching
events that had been reported in their MAs on the past year: from low,
with less than 20 events reported in the last year, to very high, with
more than 100 events reported in the year. All the variables per MA are
included in the supplementary information (Table S1).{~}

\begin{figure}
	\centering
	\includegraphics[width=0.9\linewidth]{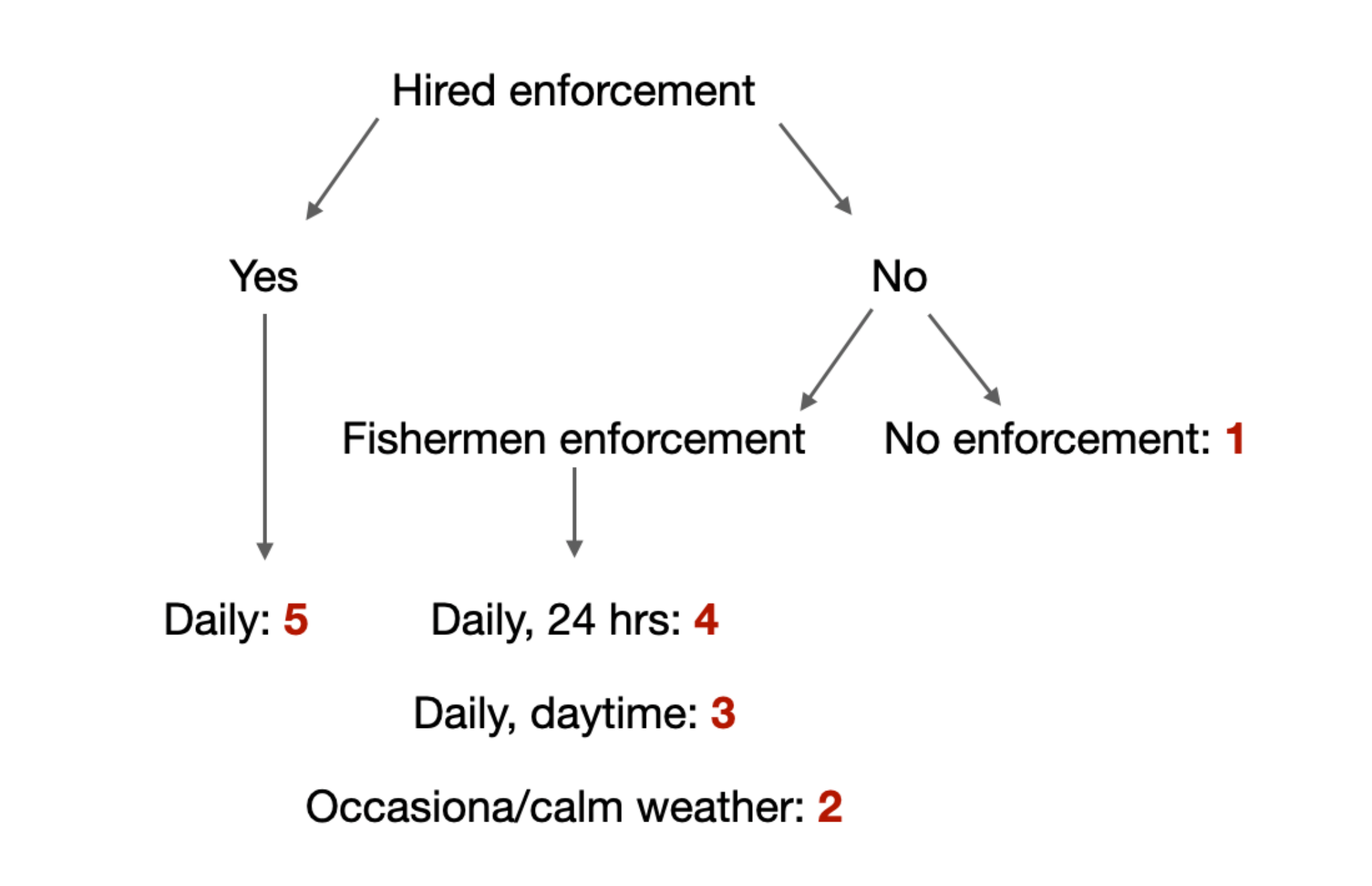}
	\caption{Ranking of enforcement level based on the fishermen's organization: payment of external personnel for the continuous enforcement (hired enforcement), the organization of the fishers to conduct the enforcement of their areas, or no enforcement of the areas. The duration of the enforcement (guard) could vary from every day, 24hrs, to only occasional. The highest enforcement is 5 (very high), lowest 1 (very low)}
	\label{fig:fig-2}
\end{figure}

\textit{Attributes of Management Areas}

Additional context variables considered relevant to assess poaching
intensity over key-hole limpet were: 1) Availability of OA area per
registered fisher in relation to the total area assigned to MA, in each
cove (index IAOA); low availability of OA per fisher due to a high
spatial density of MAs has been related with an increase of illegal
fishing in OAs due to an effort displacement {[}26{]}. Here we suggest
that such an increase in illegal fishing in OAs may yield, under certain
circumstances (like low enforcement, low compliance and/or lack of
economic alternatives), a rebound effect of increased poaching in MAs.
For each cove, the IAOA is a proxy of the proportion of OAs that
corresponds to each diver officially registered who may exert fishing
effort around that cove. It considers fishing effort density and
proportion of OA areas in relation to MA, in the accessible fishing
grounds of the cove. The lower the estimate of IAOA, the lower the
proportion of OAs in relation to MAs (less areas open to fisheries
available near a cove) and, therefore, higher poaching pressure over OAs
and over MAs as a rebound effect. Details of the methodological approach
are provided by Fernandez et al. {[}26{]}. 2) Surface of the MA, as
larger surfaces are expected to be more difficult to guard. 3) Distance
between the MAs and the official surveillance base, as this official
body is responsible for sanctioning the poachers, so fishermen contact
them when a poacher is spotted in the area; the more distant, the less
effective is enforcement as poachers must be caught in act in order to
be sanctioned. 4) Access to the MA from land through main paved routes
from the nearby location, considering also difficulty of sea access from
land, e.g., a cliff implies a difficult access; difficult access might
act as a natural protection against poachers that often reach MAs from
land. 5) Wave exposure of the fishing grounds; higher exposure acts as
natural protection, as MAs with high wave exposure are less likely to be
accessed by poachers (a high exposed area in the central coast of Chile
is an area facing south, while a protected area is facing north). 6)
Existence of alternative economic activities in the cove (e.g., tourism,
construction, recreation); this, according to fishers' leaders, results
in less poaching pressure, as fishers find alternative sources of
income. The geographical variables distance to surveillance, wave
exposure and access from land were estimated using Google Earth
software. The presence of alternative activities was depicted from the
fishers' interviews. All the variables per MA are included in the
supplementary information (Table S1).

\textit{Biological surveys}

Keyhole limpets (\emph{Fissurella} spp.) are targeted as primary
resources in the management plans of the vast majority of MAs. This
resource is currently classified as fully exploited, with minimum
landing size set at 6.5 cm for shell length in the study area, which
correspond to an age of 2 years of benthic life. In order to assess the
influence of the management regime on the proportion of undersize
limpets in the catch, a field study was conducted between October 2017
and July 2018 in MAs associated to the 13 fishers' associations (Fig.
1), as described in Fernández et al. {[}26{]}. Paired MA and OA sites
were sampled in each area by the local fishers with an observer onboard.
In each area, at least two sites were sampled since a minimum of one MA
and one OA were required. The sampling procedure was identical for the
two management regimes (MA and OA). Samples were directly obtained from
the catch of a benthic fisher, i.e., a semi-autonomous diver (hookah),
and measured onboard the fishing boat or at the landing beach or
small-scale port. Sample size is different among sites because all
individuals in the fishing bags were measured, until reaching the
minimum sample size of 200 individuals, following the protocol
established by Andreu-Cazenave et al. {[}37{]}. Size of keyhole limpets
was measured as the total length of the shell to the nearest mm.
Non-parametric Wilcoxon signed-rank tests were used to compare medians
of the size distributions of the catch between MA and OA within each
cove (W will be used to indicate the Wilcoxon test statistic). For
further detail on the analysis of the data see Fernández et al.
{[}26{]}.

\textit{Bayesian Belief Network Development}

Developed as essentially qualitative graphical models, BBNs are
especially powerful explaining the causal relationships between
variables (nodes) via conditional probability distributions. In BBNs,
the processes are not necessarily explicitly captured, on the contrary,
the expected probabilities of the outcomes are based on particular
combinations of events. Moreover, the probabilities coming from the
combination of qualitative and quantitative data can be assigned to the
BBN nodes and could come from combinations of expert opinions, empirical
field data or previous experiences cited in the literature. Since BBNs
provide a probability of an outcome rather than a discrete
(deterministic) one, a mean (expected) outcome and a confidence interval
can be determined. The way in which each input is combined to report the
probability of an outcome is determined by a weighting combination
rather than by a numerical estimation process. In other words, it is not
necessary to develop formal structural relationships linking the
different components of the model, allowing for non-linear or
discontinuous results if deemed appropriately.{~}

BBNs consist of two structural models: 1) a conceptual model (directed
acyclic graph, DAG) that represents the best available knowledge about
the functioning of the system and representing the links between the
model variables (called nodes); and 2) conditional probability tables
(CPTs) and conditional probability distributions (CPDs), which determine
the strength of the links in the DAG. Directed arrows representing
cause-effect relationships between system variables indicate the
statistical dependence between different nodes. Each arrow starts in a
parent node and ends in a child node. Feedback arrows from child nodes
to parent nodes are allowed. The DAG can be developed by experts based
on an understanding of the system and/or based on empirical
observations. This initial structure of the DAG is the basis for an
operational BBN. The probabilistic relationships between the model nodes
are specified in the CPTs and CPDs. The CPTs and CPDs can be
parameterised based on expert opinion, derived from mathematical or
logical equations, or learned from the relevant empirical data
structure. The nodes are restricted to a limited number of states that
describe the probability distribution of system variables (e.g., a node
can be discrete, with incremental or decremental states or levels, or be
continuous type). The probability distribution of each node is contained
in its CPT or CPD, and each given state of one variable is associated
with a probability between 0 and 1, so that the sum of the state values
adds up to 1 (100\%).

In our study the DAG was developed iteratively based on the system
understanding of the research team (see section 2.3 for the expected
links between variables). The first DAG was created by 12 initial nodes:
10 drivers, 1) MA surface, 2) number of MA per fishermen association, 3)
distance to surveillance authority, 4) access to MA from land, 5) wave
exposure of the MA, 6) availability of OA (IAOA), 7) alternative
activities in the cove, 8) enforcement level, 9) enforcement
effectiveness, 10) perceived poaching level; and 2 response variables,
1) the proportion of keyhole limpet below the minimum landing size
(i.e., illegal proportion), as a high proportion is an indication of
higher predisposition for poaching; and 2) the difference in median size
between the MA and adjacent OA (ê, as the normalized median MA/median
OA). Values close to 1 indicate a good state of the resource with
highest median sizes in MA compared to the paired OA area. All possible
relationships between the 12 initial nodes were considered and
quantified.~The draft DAG structure included the main components
relevant to the limpet fishery (Fig. 3).{~}

Data scoping and data analysis furthered the development of the model to
select the final set of parsimonious nodes to populate the CPTs and
CPDs. Then, Bayesian structure learning via score maximisation was
performed using the tabu search {[}38{]} in the space of the DAG. The
aim was to obtain an alternative number of possible DAGs in which all
possible combinations of the input data were compared. As a
general-purpose optimisation technique and greedy search strategy, the
tabu search employs local moves designed to affect only few local
distributions, therefore, the new DAG candidate can be scored without
recomputing the full marginal likelihood {[}38{]}. ~The team of experts
inspected the post-parameterization model, and the structure of the DAG
was accepted as plausible according to the nature and robustness of the
available data. All analyses were carried out using R language and
environment for statistical computing version 3.6.2, released 2019-12-12
(http://www.r-project.org/) and the ``bnlearn'' package v.4.6.1
{[}39{]}.

\begin{figure}
	\centering
	\includegraphics[width=0.8\linewidth]{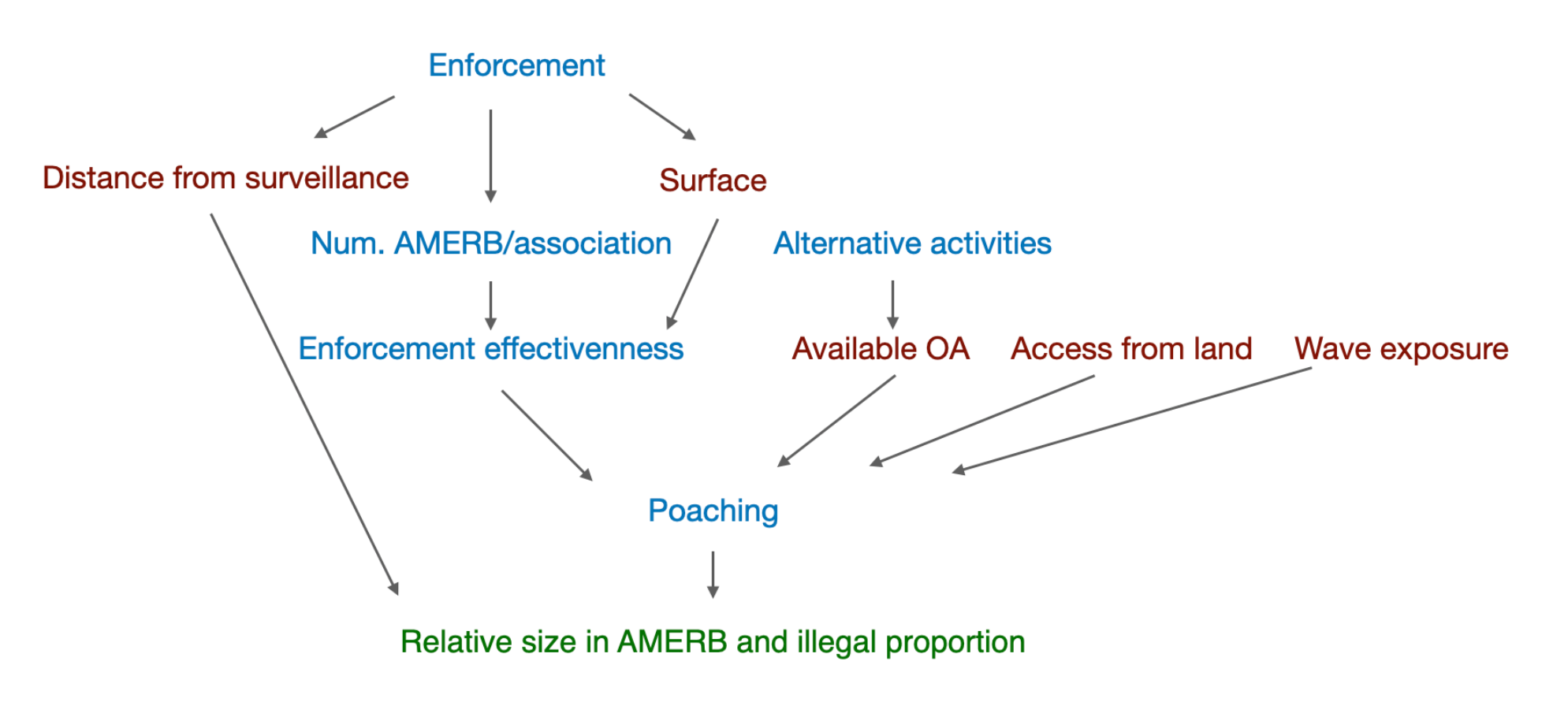}
	\caption{Possible links between variables based on expert knowledge. In red, MA attributes; in blue, variables obtained from the fishers' interviews; response (biological) variables in green.}
	\label{fig:fig-3}
\end{figure}

\textit{Conditional probability queries or ``What would happen
  if\ldots{}'' scenarios}

A query returns the probability of a specific event given some evidence.
For example, a query could be of the type "If A occurs and B does not
occur and C is greater than X and less than Y, what is the probability
that D is greater than Z?''. Based on this approach, a set of scenarios
were explored to assess the influence of the external drivers on the
state of the benthic resource. The scenarios considered that the
management target is to improve the state of keyhole limpet stock and to
reduce the proportion of limpets below minimum landing size. In our case
study, the variables that are susceptible to be managed are those linked
to the fishers' association and management bodies. The conditional
queries first consider the {probabilities of the response variable under
conditional drivers based on 2000 permutations. The conditional drivers
were enforcement level, distance to surveillance, availability of OA
area and alternative economic activities in the cove. The management
scenarios were also explored the other way around: what would be the
probabilities of different values of the selected drivers for a specific
biological response? We considered as an optimal biological response a
relative median size in the MA above 0.59 and an illegal proportion of
limpet bellow 0.31. These responses occurred in \emph{ca.} 15\% of the
MAs included in the study.{~}}

RESULTS

\textit{Size structure of keyhole limpet in Management Areas}

Significant differences in median keyhole limpet sizes were found in 22
out 24 OA-MA paired comparisons (11 out of 13 coves). Although in 19 of
22 paired comparisons median size was larger in MAs than in OAs, as
expected for a well-enforced MA, in three cases the reverse pattern was
observed, suggesting unwanted management outcomes (Algarrobo MA1,
Chungungo MA1 and Chungungo MA2). Furthermore, we found two coves with
illegality levels in MAs equal to or higher than OAs: Chigualoco, where
the MA 3 showed a proportion of undersize individuals in the catch
similar to the OA (13 and 11\%, respectively) and Chungungo, where the
MAs 1 and 2 showed significantly higher proportions of undersize
individuals (62 and 71\%, respectively, see Fig. 6) in relation to the
41\% recorded in the OA (plots for median sizes in OA and MA and
statistical test are included in supplementary information; Table S2).

\textit{Enforcement level and effectiveness}

From the 39 coves where enforcement was assessed (Fig. 1), 13\% had no
enforcement, 33\% only occasional enforcement, 23\% a daily 8-hours
enforcement. The rest had daily enforcement (24 h), which is performed
either by fishers (8\%) or hired personnel (23\%). The subset of 13
coves included in the study exhibited variable enforcement, from low to
very high, with no example of no enforcement (very low); therefore, the
enforcement variable included in the BBN has 4 levels. Three fishermen
associations had low enforcement, 1 moderate, 2 high and 7 very high
enforcement. However, by considering the effectiveness of the
enforcement level endured by the fishermen association, in 20 MAs the
enforcement level was reduced by 1 or 2 scores. In 11 MAs from
associations with very high enforcement level, the effective enforcement
was lower than the formal enforcement: 1 MA changed from very high to
low; 5 from very high to moderate; 5 from very high to high
(supplementary information; Table S2).

There was no linear relationship between the level of enforcement and
the number of poaching events per year reported by the fishermen leaders
during the interviews. Generally, sites with high enforcement have
moderate-low perceived poaching; and low enforcement is related with
high or very high perceived poaching. However, in some cases, high
enforcement is related with very high perceived poaching (Quintay and
Hornos) or low enforcement is related with low perceived poaching
(Chungungo) (Fig. 4).{~}

Similarly, there is no evident relationship between the state of the
resource, measured either as the illegal proportion of keyhole limpets
or as the relative size of individuals in the catch in MAs, and the
enforcement level and enforcement effectiveness in each MA (Fig. 5). The
absence of a direct link suggests other variables are playing a role in
the state of the benthic resource.{~}

\begin{figure}
	\centering
	\includegraphics[width=0.9\linewidth]{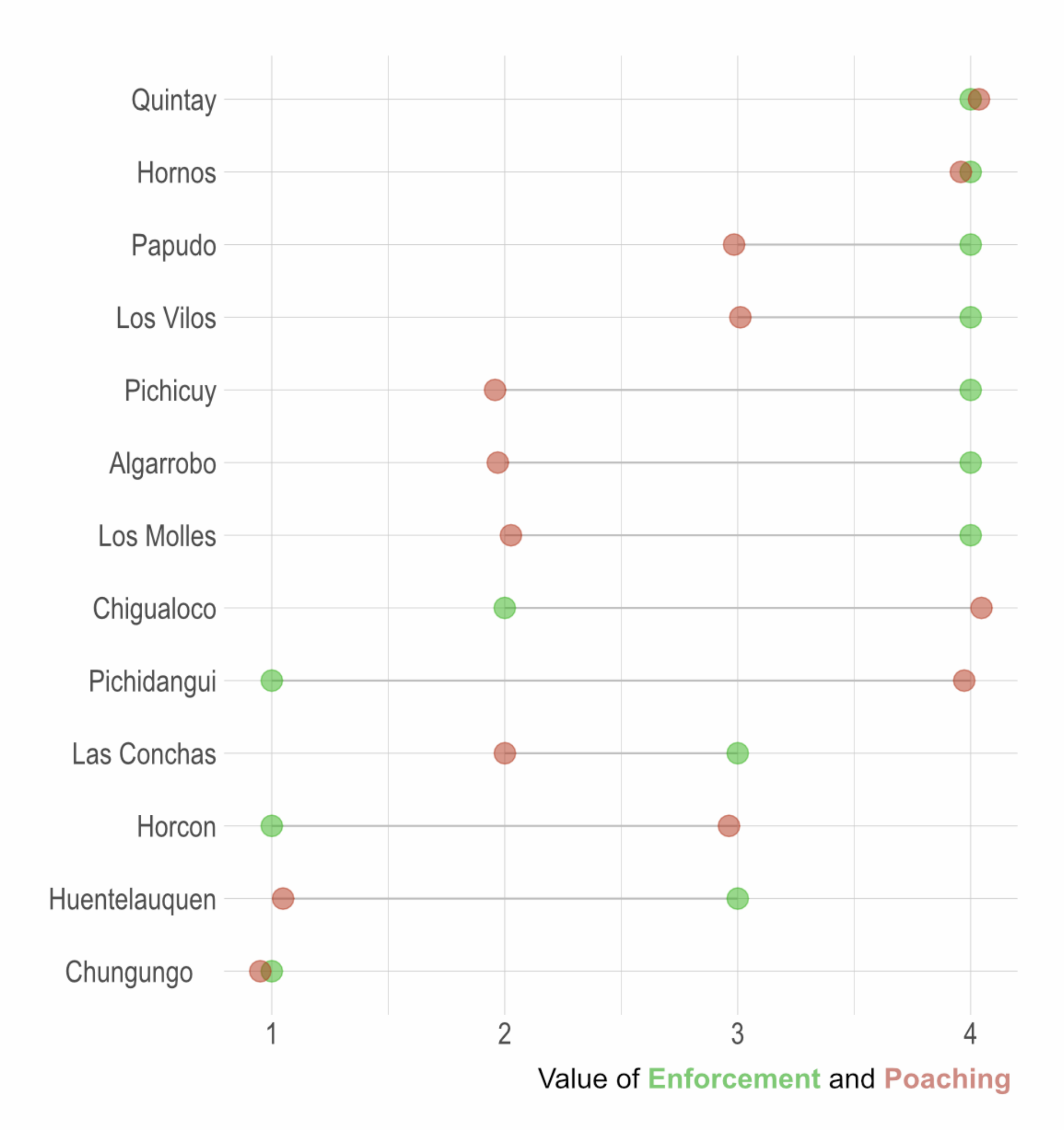}
	\caption{Cleveland dot plot showing the overlap between the ranking of enforcement endured by the fishers' association (green dots, 1, low, to 4, high) and the level of poaching perceived by the fishermen (red dots, 1, low, to 4, high)}
	\label{fig:fig-4}
\end{figure}

\begin{figure}
	\centering
	\includegraphics[width=0.9\linewidth]{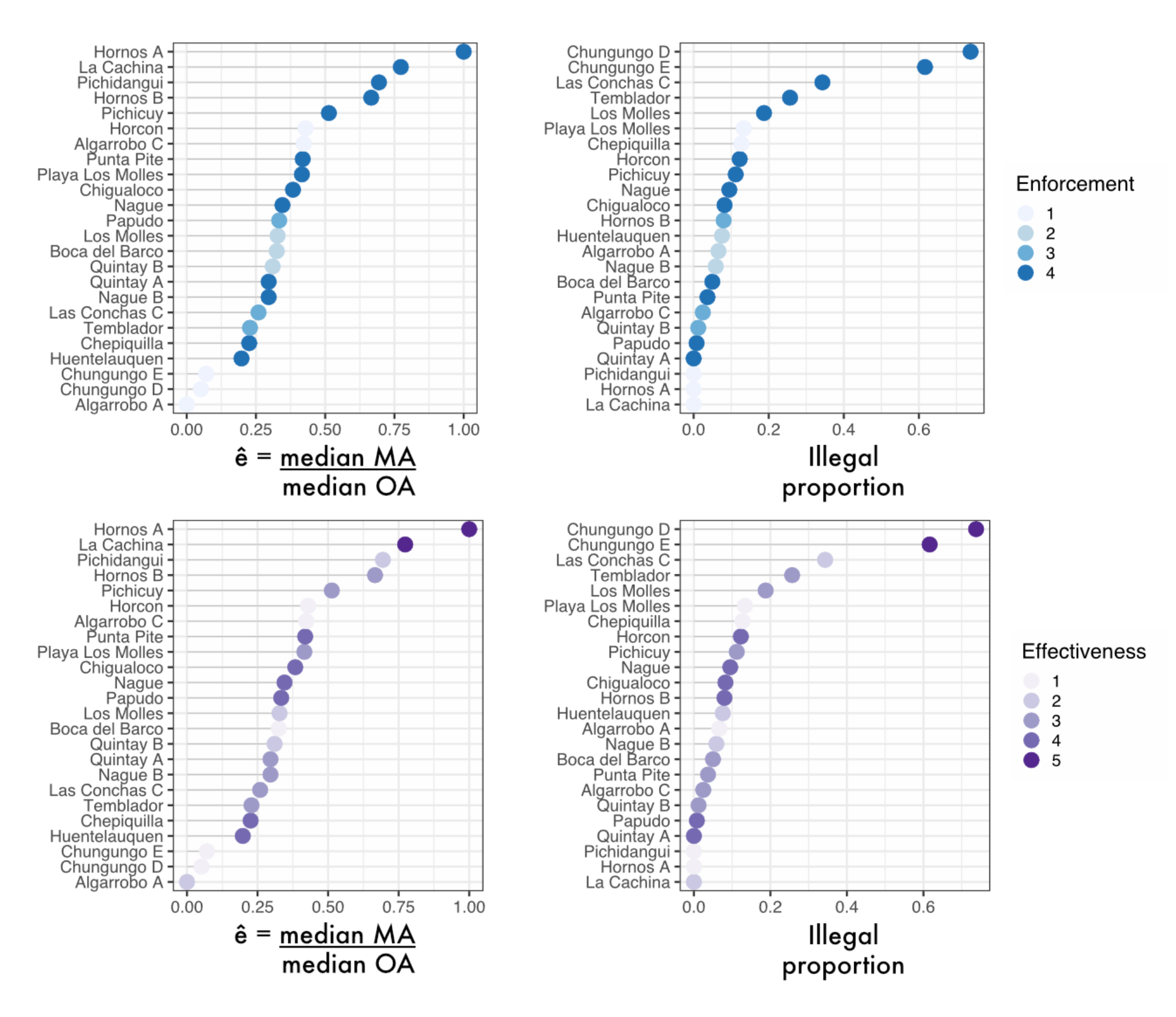}
	\caption{Relationship between the relative median size of individual limpets fished in the MA (ê), left panels, and the proportion of illegal limpets in the catch, right panels, and the enforcement level (upper panels) and enforcement effectiveness (bottom panels) in the case studies (24 MAs, indicated by their local names).}
	\label{fig:fig-5}
\end{figure}

\textit{Bayesian-belief network}

A BBN approach was adopted to explore the influence of a set of external
potential drivers of poaching intensity over keyhole limpet state within
the MAs (Fig. 3). Keyhole limpet state was assessed through the relative
size of limpets fished in MAs and the illegal proportion of limpets in
the catch (see methods for further details). The limpet state is
considered a proxy for the effects of poaching on benthic resources. The
level of poaching was also assessed relying on fishermen perceptions
(supplementary information; Table S1); however, the perceived poaching
exhibited an unexpected link with the condition of the resource, as
lower reported poaching events were linked with lower average sizes in
MA and higher proportions of illegal size. Therefore, this variable was
excluded from the final BBN. Access and number of MAs were also excluded
from the optimal network, as these variables had no link with the
biological variables in our case study.{~}

The most parsimonious BBN included most links that were predicted by
expert knowledge in the draft DAG (Fig. 3 \emph{vs}. Fig. 6). However,
some links were rather unexpected (Fig. 6). For example, the existence
of alternative activities plays a major role in the network, influencing
both the magnitude of the effects of availability of OA area (with less
fishing pressure rebounding on the MAs) and enforcement level, as
alternative activities might reduce the time dedicated to survey the
MAs. On the other hand, availability of OA area is linked to the illegal
proportion of limpet through the distance to the surveillance authority
and, therefore, poaching reports are not effective. Enforcement has
effects over the relative size in the MA (ê) through the effectiveness
of enforcement; therefore, it highlights the relevance of uneven
surveillance efforts across several MAs controlled by a single
association. Some of the links were probably casual, considering the
limited number of coves and MAs included in the study (Fig. 6):
availability of OAs and distance to surveillance is linked to the
surface of the MA and to wave exposure. Whereas the availability of OAs
is controlled by both the number and extension of the MA available per
fisher in an area of the coast, probably the sections of the coast most
exposed to waves, and/or less accessible from the coves, tend to
concentrate less MAs.{~}

\begin{figure}
	\centering
	\includegraphics[width=0.9\linewidth]{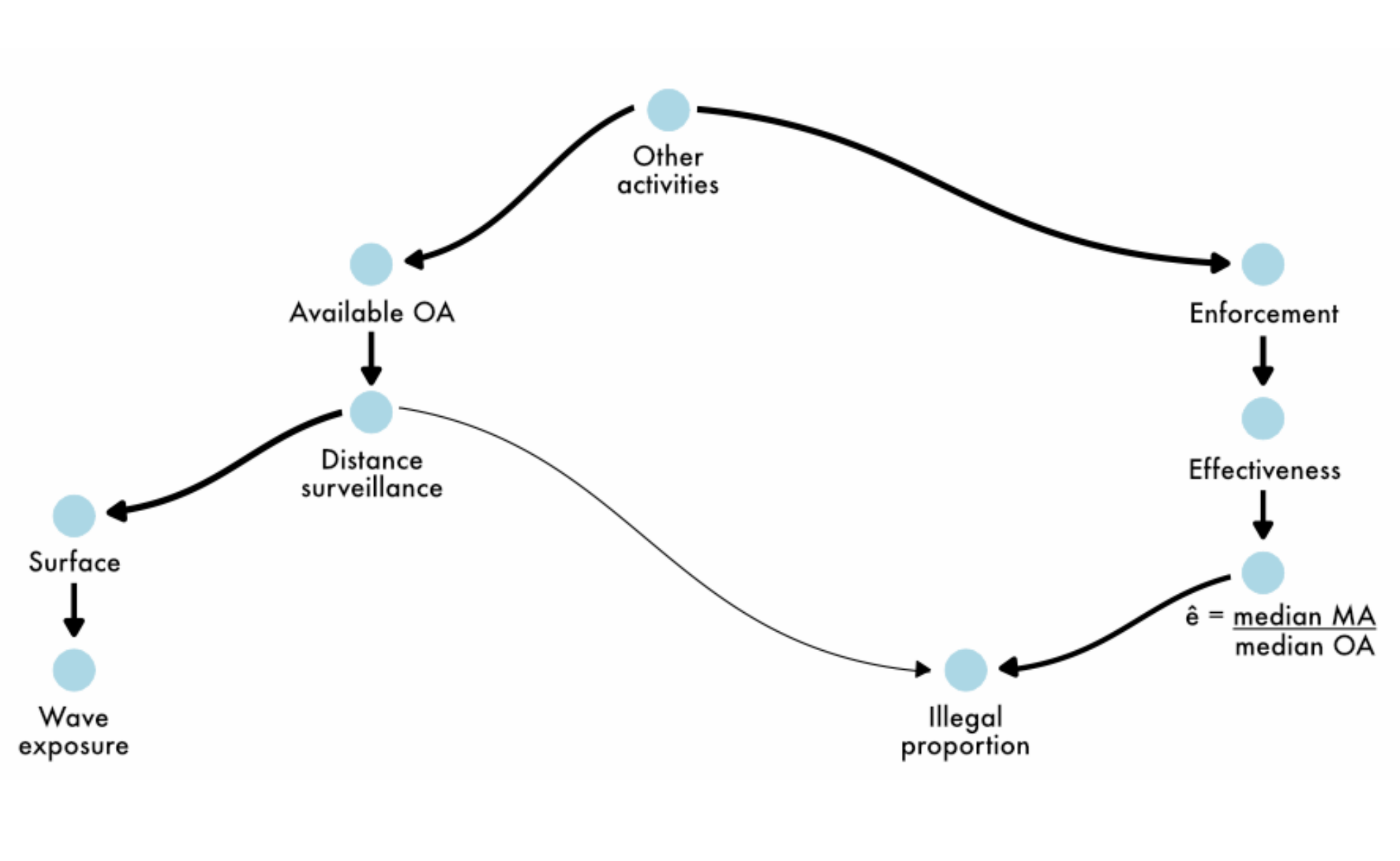}
	\caption{Bayesian-belief Network obtained for the case study. The thickness of the arrow represents the strength of the link.}
	\label{fig:fig-6}
\end{figure}

\textit{Scenarios to achieve a good state of the fishery resource}

In order to observe a low proportion of illegal limpets in the catch,
the ideal scenario includes a high availability of OA areas per fisher
and short distance to a surveillance authority base (scenario 2 in Table
1). The combination of any of these variables with alternative economic
activities for the fishers and high enforcement increases the
probability of a lower proportion of limpets of illegal size in the
catch (scenario 1, 3 and in Table 1). In particular, the combination of
high availability of OA areas with any of the other three variables
yielded a high probability (\textgreater{}0.88) of an illegal proportion
below 30\% of the catch. On the other hand, the external drivers had a
weak effect on the relative median size of limpets fished in MAs, with
values closer to 1 indicating larger median sizes in the MA compared to
neighbouring OA. Either high enforcement or high effectiveness combined
with the presence of other activities did not yield probabilities higher
than 39\% (Table 1).{~}

{Table 1. Queries performed over the BBN (Fig. 6). Probabilities
of the response variable under conditional drivers based on 2000
permutations}

\begin{longtable}[]{@{}lllll@{}}
\toprule
\begin{minipage}[t]{0.17\columnwidth}\raggedright
\textbf{Scenarios}\strut
\end{minipage} & \begin{minipage}[t]{0.17\columnwidth}\raggedright
\textbf{Variable }1\strut
\end{minipage} & \begin{minipage}[t]{0.17\columnwidth}\raggedright
\textbf{Variable 2}\strut
\end{minipage} & \begin{minipage}[t]{0.17\columnwidth}\raggedright
\textbf{Response}\strut
\end{minipage} & \begin{minipage}[t]{0.17\columnwidth}\raggedright
\textbf{Probability}\strut
\end{minipage}\tabularnewline
\begin{minipage}[t]{0.17\columnwidth}\raggedright
Sce. 1\strut
\end{minipage} & \begin{minipage}[t]{0.17\columnwidth}\raggedright
Available OA = very high or high\strut
\end{minipage} & \begin{minipage}[t]{0.17\columnwidth}\raggedright
Other activities = Y\strut
\end{minipage} & \begin{minipage}[t]{0.17\columnwidth}\raggedright
Illegal proportion \textless{}= 0.3\strut
\end{minipage} & \begin{minipage}[t]{0.17\columnwidth}\raggedright
0.881\strut
\end{minipage}\tabularnewline
\begin{minipage}[t]{0.17\columnwidth}\raggedright
Sce. 2\strut
\end{minipage} & \begin{minipage}[t]{0.17\columnwidth}\raggedright
Available OA =very high or high\strut
\end{minipage} & \begin{minipage}[t]{0.17\columnwidth}\raggedright
Distance to surveillance = close\strut
\end{minipage} & \begin{minipage}[t]{0.17\columnwidth}\raggedright
Illegal proportion \textless{}= 0.3\strut
\end{minipage} & \begin{minipage}[t]{0.17\columnwidth}\raggedright
0.935\strut
\end{minipage}\tabularnewline
\begin{minipage}[t]{0.17\columnwidth}\raggedright
Sce. 3\strut
\end{minipage} & \begin{minipage}[t]{0.17\columnwidth}\raggedright
Available OA =very high or high\strut
\end{minipage} & \begin{minipage}[t]{0.17\columnwidth}\raggedright
Enforcement = very high or high\strut
\end{minipage} & \begin{minipage}[t]{0.17\columnwidth}\raggedright
Illegal proportion \textless{}= 0.3\strut
\end{minipage} & \begin{minipage}[t]{0.17\columnwidth}\raggedright
0.961\strut
\end{minipage}\tabularnewline
\begin{minipage}[t]{0.17\columnwidth}\raggedright
Sce. 4\strut
\end{minipage} & \begin{minipage}[t]{0.17\columnwidth}\raggedright
Enforcement = very high or high\strut
\end{minipage} & \begin{minipage}[t]{0.17\columnwidth}\raggedright
Other activities = Y\strut
\end{minipage} & \begin{minipage}[t]{0.17\columnwidth}\raggedright
Relative size at MA \textgreater{}= 0.4\strut
\end{minipage} & \begin{minipage}[t]{0.17\columnwidth}\raggedright
0.378\strut
\end{minipage}\tabularnewline
\begin{minipage}[t]{0.17\columnwidth}\raggedright
Sce. 5\strut
\end{minipage} & \begin{minipage}[t]{0.17\columnwidth}\raggedright
Effectiveness = moderate to very high{~}\strut
\end{minipage} & \begin{minipage}[t]{0.17\columnwidth}\raggedright
Other activities = Y{~}\strut
\end{minipage} & \begin{minipage}[t]{0.17\columnwidth}\raggedright
Relative size at MA \textgreater{}= 0.6\strut
\end{minipage} & \begin{minipage}[t]{0.17\columnwidth}\raggedright
0.394\strut
\end{minipage}\tabularnewline
\begin{minipage}[t]{0.17\columnwidth}\raggedright
Sce. 6\strut
\end{minipage} & \begin{minipage}[t]{0.17\columnwidth}\raggedright
Other activities = Y\strut
\end{minipage} & \begin{minipage}[t]{0.17\columnwidth}\raggedright
Distance to surveillance = close\strut
\end{minipage} & \begin{minipage}[t]{0.17\columnwidth}\raggedright
Illegal proportion \textless{}= 0.3\strut
\end{minipage} & \begin{minipage}[t]{0.17\columnwidth}\raggedright
0.991\strut
\end{minipage}\tabularnewline
\begin{minipage}[t]{0.17\columnwidth}\raggedright
Sce. 7\strut
\end{minipage} & \begin{minipage}[t]{0.17\columnwidth}\raggedright
Effectiveness = very high or high\strut
\end{minipage} & \begin{minipage}[t]{0.17\columnwidth}\raggedright
Distance to surveillance = close\strut
\end{minipage} & \begin{minipage}[t]{0.17\columnwidth}\raggedright
Illegal proportion \textless{}= 0.3\strut
\end{minipage} & \begin{minipage}[t]{0.17\columnwidth}\raggedright
0.987\strut
\end{minipage}\tabularnewline
\bottomrule
\end{longtable}

In order to attain a good state of the fishery resource, with an illegal
proportion of limpet catch below 30\% and the relative median size in
the MA over 60\% (i.e., larger limpets in the MA in reference to
neighbouring OA), the optimal combination of variables indicate that 1)
availability of OA area should be high to very high (probability of
0.53), 2) the enforcement should be high to very high (probability of
0.80), 3) the effectiveness can be maintained at moderate (probability
of 0.77), 4) the distance to surveillance should be short (probability
of 0.59), 4) while the presence or absence of other activities does not
have a major effect (Table 2).{~}

\begin{center}
	\includegraphics[width=1\linewidth]{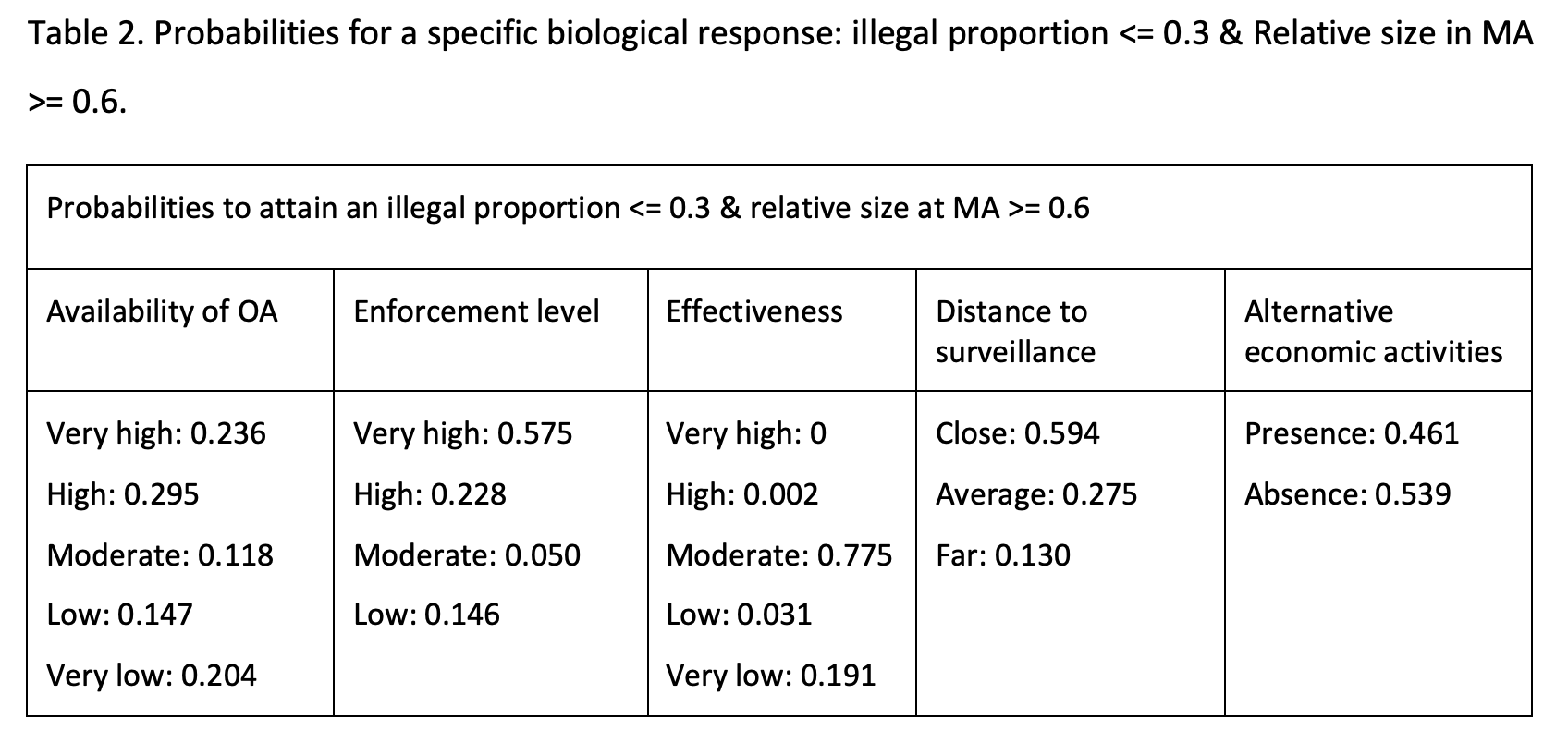}
\end{center}

DISCUSSION

To disentangle SSF complex dynamics, innovative approaches are necessary
to deal with the scarcity of data and the need to take advantage of both
scientific expertise and fishers' knowledge {[}4{]}. Bayesian-belief
networks are useful for SSF research due to flexibility in incorporating
data of different nature, where expert knowledge plays a key role. The
BBN exercise on a TURF system in central Chile allowed identifying the
role of multiple external socio-economic and geographical context
drivers on the sustainability of the fishery resource. Our BBN provided
evidence of the absence of a direct link between the level of MA
enforcement and the state of the benthic resource, with other
socio-economic (e.g., alternative economic activities) and context
variables (e.g., fishing effort pressure or distance to surveillance
authorities) playing important roles. The interviews with fishers
evidenced high variability in enforcement protocols and also that a
concern frequently shared by the fishers' leaders was the high cost of
allocating human resources to survey the area, as these costs must be
covered by the fishers' association. Often, one association is in charge
of more than one MA, which prompts the allocation of the limited
enforcement resources only to one area, either the most accessible or
the most productive one. \textbf{} This variability was considered in
the conception of the variable ``effective enforcement'' and the model
evidenced the importance of this variable in the state of the resource,
as in some cases the level of enforcement achieved by the fishermen
association was lowered by an uneven and non-effective enforcement
across MAs. These observations are aligned with previous results in the
same study area where fishers reported that their organizations often
decide not to monitor the MAs that are furthest away (less accessible)
from the cove {[}23{]}.{~}

The relationship between the enforcement level and the number of
poaching events per year reported by the fishers was not consistent,
although an opposite relationship was observed in several coves. The
observed lack of consistency in these results might be showing a bias
due to the negative fishers' perception on the effectiveness of the
poaching sanctioning system {[}23{]}. Lab-in-field experiments focusing
on the effect of co-enforcement on limiting the access to common pool
resources in the study area showed that the existence of external
poaching sanctions (like those imposed by a government authority)
improved the willingness of resource users to invest in enforcement,
ending up with a reduction in poaching {[}40{]}. Compliance-monitoring
agencies in Chile concentrate on access control with much less emphasis
on compliance to quotas, bans or minimum legal size. According to the
national fisheries law, the National Fisheries Service SERNAPESCA has to
publicly deliver an annual report of enforcement actions, which, in its
last edition reveals the low percentage of enforcement work achieved in
the central coast of Chile (8.8\% of the total field enforcement actions
in the country), where the present study focuses {[}41{]}. This might be
related with the geographic location of coves, several of them very
distant from the closest city, where enforcement agencies are usually
based. The implementation of local enforcement tools must be necessarily
considered to address this difficulty.{~ }The sustainability of SSF is a
complex socio-economic and ecological problem {[}42{]}, with many local
variables playing an important role and no bullet-proof solution. In
fact, illegal fishing of Chilean SSF has been termed as a wicked problem
by Nahuelhual et al. {[}43{]}, as it is characterised by its complexity,
uncertainty and interdependence of factors. Therefore, a
multi-dimensional assessment, such as the one presented here, is
necessary to address the illegal fishing of benthic resources in
Chile.{~}

A Bayesian-belief network is a strong tool to address multi-dimensional
problems, as it nourishes from expert knowledge and it is flexible to be
used for scenario analysis by local actors. In our study, the optimal
model included most of the links predicted by expert knowledge (Fig. 3
\emph{vs}. Fig. 7). However, some links were rather unexpected. For
example, the existence of alternative economic activities is probably
alleviating fishing pressure around the coves, and, in the case of areas
with low availability of OAs, this translates into lower poaching in the
MAs. The scenario analysis explores variables that are susceptible to be
managed, including distance to surveillance authorities and a spatial
planning approach to fisheries restrictions. It is legitimate to state
that, according to our model, lowering illegality in benthic artisanal
fisheries in central Chile depends mostly on two government
administrative matters: i) improving authority surveillance mechanisms
and rules (in order to increase effectiveness of anti-poaching controls
and reduce the negative consequences of higher distances to surveillance
authority bases); ii) specific actions in marine spatial planning of the
TURF system (in order to increase the availability of OAs around each
MAs).{~}

An additional administrative issue is the closure of the delivery of
artisanal fishing permits for more than 10 years, that has let young
generations of fishers with no choice but to fish illegally {[}43{]}.
Fishers leaders' perceptions gathered during this study support this
statement, as leaders often pointed out that there is an increasingly
large proportion of young, un-associated, fishers that are discouraged
by the current co-management system that does not fulfil the
cost-benefit balance. The recurrent idea among different stakeholders
(i.e., fishers, government and intermediaries of resource value chain)
is that illegal fishing is a consequence of the lack of opportunities
and economic needs {[}43,44{]}, reflected by the variable ``alternative
economic activities'' in our model. Thus, managers must ensure fishers
do not have ``a reason to poach'', by putting efforts on the
socio-economic well-being of the cove. But, bearing in mind the later as
a necessary long-term task, immediate actions on preventing poaching
must be considered. Unmonitored and unsanctioned poaching in TURFs may
impede fishers of sustaining their resources and benefit from harvesting
surplus. Contrastingly, higher sanctions to poaching may both help
fishers to defend their TURF resources and improve un-associated
fishers' willingness to bind to a TURF system {[}40{]}. Nevertheless, a
strengthening in the sanctions to poaching must be framed in the
artisanal SSF perspective {[}12{]}, where non-compliance is strongly
conditioned by local socio-economic contexts, as mentioned above
{[}14{]}. Currently, fishers must cover the (increasingly high) cost of
surveillance of their TURFs but rely on the administration to punish
poachers {[}24{]}. Although the legislation considers formal sanctions
to poaching and illegal practices, these are informally allowed by the
enforcement agencies. This suggests a vicious circle around the problem
of poaching in TURFs: economic needs prompt poaching and deter fishers
to invest in surveillance, in response the government administration
strengthens formal sanctions but does not improve enforcement
mechanisms, letting the SSF stuck in an illegality trap {[}8{]}.

From our results, we can suggest that illegal fishing on key-hole limpet
in central Chile arises from a complex network of drivers, from
methodological to logistic problems, which appear to be linked to
monitoring of compliance, lack of support of government agencies in the
co-management process, socio-economic context of fishers, and an
unsuitable spatial planning of the TURFs. These results are in line with
observations focusing in other benthic resources like the highly
valuable loco or Chilean abalone {[}27{]} and the king crab {[}43{]}. A
change in paradigm to a local perspective in the implementation of
mechanisms to overcome enforcement and planning weaknesses is needed.
This change in paradigm should be aided by suitable diagnosis of local
socio-economic contexts of fishers, and the improvement of local and
regional governance tools (e.g., Management Committees, which are
already considered in the national regulation).{~}

In the current TURF system in Chile, fishers select an area, and it is
assigned under request, but there is no advice from decision makers and
scientists on where to allocate the area (only on quotas). There is an
inherent problem on the planification of the system that might lead, for
example, to high fishing pressure in adjacent open access areas
{[}22,26{]}. Ospina-Alvarez et al. {[}45{]} showed the benefits of a
planned network of MAs in the central coast of chile that would have
positive effects on both fisheries and biodiversity conservation.
However, this spatial modelling exercise also evidenced the importance
of enforcement for the effective functioning of the restricted areas
network and the need to consider all environmental and human factors in
the optimization criteria. High fishing pressure on the OA areas, with
overexploited benthic resources {[}37{]}, has a negative feedback on the
MAs that are exposed to illegal extraction of a benthic resource that is
generally in a better condition than neighbouring OA areas {[}46{]}. A
systematic and science-based spatial planning of the fishery restricted
areas would avoid high concentrations of MAs in sections of the coast,
generally those more densely populated, while identifying the most
productive areas for the placement of fishery restricted areas {[}45{]}.
This spatial approach to the problem would alleviate fishing pressure on
the restricted MAs; however, additional actions, like increasing
surveillance efforts, are necessary for the effective performance of a
network of well-enforced MA.{~}

While TURFs systems aim to incentive fishers' care of their fishing
grounds, these need to be well planned and supported financially and
logistically by the administration. Property rights schemes could
benefit from a re-formulation that considers the social rights and
benefits (beyond the individual rights) and fishers' traditional
ecological knowledge {[}6{]}, so the conservation of the ecosystem is a
priority over the individual interests to exploit the resources
{[}47{]}. However, in the case study, the top-down component of the
co-management system seems to fail as there is poor planning and
financing. The decision-making power granted to fishers in the current
co-management process is very much limited to a MA monitoring and
surveillance, with insufficient resources and support to carry out this
task. In Chile, and in SSF in general, the balance of the
fishermen-administration role must be planned to ensure a fully
democratised system that is able to cope with illegal fishing and
overexploitation. Addressing these complex problems inherent to the
functioning of SSF are central, as SSF can play a key role in the
sustainable development of global fisheries.{~}

\textbf{Acknowledgements}

This work was partly funded by a Fondecyt Project 1171603 grant to MDS
and MF and the Iniciativa Científica Milenio (Project CCM RC 1300004)
from Ministerio de Economía, Fomento y Turismo de Chile. The authors
want to{ thank all participants of the survey}s{ for their essential
contribution to this study.{~}}

\textbf{Author contribution}

Silvia de Juan: conceptualization, investigation, methodology, formal
analysis, writing-original draft. Dulce Subida: conceptualization,
investigation, formal analysis, writing-original draft. Andres
Ospina-Alvarez: methodology, software, formal analysis. Ainara Aguilar:
investigation. Miriam Fernandez: conceptualization, writing-review and
editing, funding acquisition.

\textbf{References}

{[}1{]}{ }J. Jacquet, D. Pauly, Funding Priorities: Big Barriers to
Small-Scale Fisheries: \emph{Funding for Fisheries}, Conservation
Biology. 22 (2008) 832--835.
https://doi.org/10.1111/j.1523-1739.2008.00978.x.

{[}2{]}{ }FAO, Securing sustainable small-scale fisheries: sharing good
practices from around the world, FAO, 2019.

{[}3{]}{ }C. Costello, D. Ovando, R. Hilborn, S.D. Gaines, O. Deschenes,
S.E. Lester, Status and Solutions for the World's Unassessed Fisheries,
Science. 338 (2012) 517--520. https://doi.org/10.1126/science.1223389.

{[}4{]}{ }C. Pita, S. Villasante, J.J. Pascual-Fernández, Managing
small-scale fisheries under data poor scenarios: lessons from around the
world, Marine Policy. 101 (2019) 154--157.
https://doi.org/10.1016/j.marpol.2019.02.008.

{[}5{]}{ }M.-J. Roux, R.F. Tallman, Z.A. Martin, Small-scale fisheries
in Canada's Arctic: Combining science and fishers knowledge towards
sustainable management, Marine Policy. 101 (2019) 177--186.
https://doi.org/10.1016/j.marpol.2018.01.016.

{[}6{]}{ }C. Ruano-Chamorro, M.D. Subida, M. Fernandez, Fishers'
perception: An alternative source of information to assess the data-poor
benthic small-scale artisanal fisheries of central Chile, Ocean and
Coastal Management. 146 (2017) 67--76.
https://doi.org/10.1016/j.ocecoaman.2017.06.007.

{[}7{]}{ }J. Aburto, W. Stotz, Learning about TURFs and natural
variability: Failure of surf clam management in Chile, Ocean \& Coastal
Management. 71 (2013) 88--98.
https://doi.org/10.1016/j.ocecoaman.2012.10.013.

{[}8{]}{ }L. Nahuelhual, G. Saavedra, M.A. Mellado, X.V. Vergara, T.
Vallejos, A social-ecological trap perspective to explain the emergence
and persistence of illegal fishing in small-scale fisheries, Maritime
Studies. 19 (2020) 105--117. https://doi.org/10.1007/s40152-019-00154-1.

{[}9{]}{ }B. Worm, R. Hilborn, J.K. Baum, T.A. Branch, J.S. Collie, C.
Costello, M.J. Fogarty, E.A. Fulton, J.A. Hutchings, S.R. Jennings, O.P.
Jensen, H.K. Lotze, P.M. Mace, T.R. McClanahan, C. Minto, S.R. Palumbi,
A.M. Parma, D. Ricard, A.A. Rosenberg, R. Watson, D. Zeller, Rebuilding
global fisheries, Science. 325 (2009) 578--585.

{[}10{]}{ }C. Mora, R.A. Myers, M. Coll, S. Libralato, T.J. Pitcher, R.
Sumaila, D. Zeller, R. Watson, K.J. Gaston, B. Worm, Management
effectiveness of the world's marine fisheries, PLoS Biology. 7 (2009)
e1000131. https://doi.org/10.1371/journal.pbio.1000131.

{[}11{]}{ }D.J. Agnew, J. Pearce, G. Pramod, T. Peatman, R. Watson, J.R.
Beddington, T.J. Pitcher, Estimating the Worldwide Extent of Illegal
Fishing, PLoS ONE. 4 (2009) e4570.
https://doi.org/10.1371/journal.pone.0004570.

{[}12{]}{ }C.J. Donlan, C. Wilcox, G.M. Luque, S. Gelcich, Estimating
illegal fishing from enforcement officers, Sci Rep. 10 (2020) 12478.
https://doi.org/10.1038/s41598-020-69311-5.

{[}13{]}{ }P. Veiga, C. Pita, M. Rangel, J.M.S. Gonçalves, A. Campos,
P.G. Fernandes, A. Sala, M. Virgili, A. Lucchetti, J. Brčić, S.
Villasante, M.A. Ballesteros, R. Chapela, J.L. Santiago, S. Agnarsson,
Ó. Ögmundarson, K. Erzini, The EU landing obligation and European
small-scale fisheries: What are the odds for success?, Marine Policy. 64
(2016) 64--71. https://doi.org/10.1016/j.marpol.2015.11.008.

{[}14{]}{ }R. Oyanedel, S. Gelcich, E.J. Milner‐Gulland, Motivations for
(non‐)compliance with conservation rules by small‐scale resource users,
CONSERVATION LETTERS. (2020). https://doi.org/10.1111/conl.12725.

{[}15{]}{ }E. Ostrom, V. Ostrom, Public economy organization and service
delivery, Dearborn University of Michigan, 1977.
http://dlc.dlib.indiana.edu/dlc/handle/
10535/732.

{[}16{]}{ }W. Battista, R. Romero-Canyas, S.L. Smith, J. Fraire, M.
Effron, D. Larson-Konar, R. Fujita, Behavior Change Interventions to
Reduce Illegal Fishing, Front. Mar. Sci. 5 (2018) 403.
https://doi.org/10.3389/fmars.2018.00403.

{[}17{]}{ }C. Nguyen Thi Quynh, S. Schilizzi, A. Hailu, S. Iftekhar,
Territorial Use Rights for Fisheries (TURFs): State of the art and the
road ahead, Marine Policy. 75 (2017) 41--52.
https://doi.org/10.1016/j.marpol.2016.10.004.

{[}18{]}{ }M. Fernández, J.C. Castilla, Marine Conservation in Chile:
Historical Perspective, Lessons, and Challenges, Conservation Biology.
19 (2005) 1752--1762. https://doi.org/10.1111/j.1523-1739.2005.00277.x.

{[}19{]}{ }J. Aburto, G. Gallardo, W. Stotz, C. Cerda, C.
Mondaca-Schachermayer, K. Vera, Territorial user rights for artisanal
fisheries in Chile -- intended and unintended outcomes, Ocean \& Coastal
Management. 71 (2013) 284--295.
https://doi.org/10.1016/j.ocecoaman.2012.09.015.

{[}20{]}{ }S. Gelcich, M. Fernandez, N. Godoy, A. Canepa, L. Prado, J.C.
Castilla, Territorial user rights for fisheries as ancillary instruments
for marine coastal conservation in Chile., Conservation Biology. 26
(2012) 1005--1015. https://doi.org/10.1111/j.1523-1739.2012.01928.x.

{[}21{]}{ }A. Pérez-Matus, A. Ospina-Alvarez, P.A. Camus, S.A. Carrasco,
M. Fernandez, S. Gelcich, N. Godoy, F.P. Ojeda, L.M. Pardo, N.
Rozbaczylo, M.D. Subida, M. Thiel, E.A. Wieters, S.A. Navarrete,
Temperate rocky subtidal reef community reveals human impacts across the
entire food web, Marine Ecology Progress Series. 567 (2017) 1--16.
https://doi.org/10.3354/meps12057.

{[}22{]}{ }J. Beckensteiner, A.M. Scheld, M. Fernández, D.M. Kaplan,
Drivers and trends in catch of benthic resources in Chilean TURFs and
surrounding open access areas, Ocean \& Coastal Management. 183 (2020)
104961. https://doi.org/10.1016/j.ocecoaman.2019.104961.

{[}23{]}{ }K.J. Davis, M.E. Kragt, S. Gelcich, M. Burton, S. Schilizzi,
D.J. Pannell, Why are Fishers not Enforcing Their Marine User Rights?,
Environ Resource Econ. 67 (2017) 661--681.
https://doi.org/10.1007/s10640-015-9992-z.

{[}24{]}{ }S. Gelcich, J. Cinner, C. Donlan, S. Tapia-Lewin, N. Godoy,
J. Castilla, Fishers' perceptions on the Chilean coastal TURF system
after two decades: problems, benefits, and emerging needs, Bms. 93
(2017) 53--67. https://doi.org/10.5343/bms.2015.1082.

{[}25{]}{ }R.M. Bandin Llanos, Impacto de la captura ilegal en la
pesquería del recurso 'loco' (Concholepas concholepas) En áreas de
manejo y explotación de recursos bentónicos: El caso de la Isla Mocha,
Thesis, Universidad de Concepcion, 2013.
http://repositorio.udec.cl/jspui/handle/11594/763.

{[}26{]}{ }M. Fernández, M. Kriegl, V. Garmendia, A. Aguilar, M.D.
Subida, Evidence of illegal catch in the benthic artisanal fisheries of
central Chile: patterns across species and management regimes, Lat. Am.
J. Aquat. Res. 48 (2020) 287--303.
https://doi.org/10.3856/vol48-issue2-fulltext-2475.

{[}27{]}{ }R. Oyanedel, A. Keim, J.C. Castilla, S. Gelcich, Illegal
fishing and territorial user rights in Chile: Illegal Fishing,
Conservation Biology. 32 (2018) 619--627.
https://doi.org/10.1111/cobi.13048.

{[}28{]}{ }H.M. Ballesteros, G. Rodríguez-Rodríguez, ``Acceptable'' and
``unacceptable'' poachers: Lessons in managing poaching from the
Galician shellfish sector, Marine Policy. 87 (2018) 104--110.
https://doi.org/10.1016/j.marpol.2017.10.015.

{[}29{]}{ }L. Little, S. Kuikka, A. Punt, F. Pantus, C. Davies, B.
Mapstone, Information flow among fishing vessels modelled using a
Bayesian network, Environmental Modelling \& Software. 19 (2004) 27--34.
https://doi.org/10.1016/S1364-8152(03)00100-2.

{[}30{]}{ }P. Haapasaari, C.G.J. Michielsens, T.P. Karjalainen, K.
Reinikainen, S. Kuikka, Management measures and fishers' commitment to
sustainable exploitation: a case study of Atlantic salmon fisheries in
the Baltic Sea, ICES Journal of Marine Science. 64 (2007) 825--833.
https://doi.org/10.1093/icesjms/fsm002.

{[}31{]}{ }P. Levontin, S. Kulmala, P. Haapasaari, S. Kuikka,
Integration of biological, economic, and sociological knowledge by
Bayesian belief networks: the interdisciplinary evaluation of potential
management plans for Baltic salmon, ICES Journal of Marine Science. 68
(2011) 632--638. https://doi.org/10.1093/icesjms/fsr004.

{[}32{]}{ }I. van Putten, A. Lalancette, P. Bayliss, D. Dennis, T.
Hutton, A. Norman-López, S. Pascoe, E. Plagányi, T. Skewes, A Bayesian
model of factors influencing indigenous participation in the Torres
Strait tropical rocklobster fishery, Marine Policy. 37 (2013) 96--105.
https://doi.org/10.1016/j.marpol.2012.04.001.

{[}33{]}{ }C.A. Pollino, O. Woodberry, A. Nicholson, K. Korb, B.T. Hart,
Parameterisation and evaluation of a Bayesian network for use in an
ecological risk assessment, Environmental Modelling \& Software. 22
(2007) 1140--1152. https://doi.org/10.1016/j.envsoft.2006.03.006.

{[}34{]}{ }E. Hoshino, I. van Putten, W. Girsang, B.P. Resosudarmo, S.
Yamazaki, A Bayesian belief network model for community-based coastal
resource management in the Kei Islands, Indonesia, E\&S. 21 (2016)
art16. https://doi.org/10.5751/ES-08285-210216.

{[}35{]}{ }M.J. Slater, Y.D. Mgaya, A.C. Mill, S.P. Rushton, S.M. Stead,
Effect of social and economic drivers on choosing aquaculture as a
coastal livelihood, Ocean \& Coastal Management. 73 (2013) 22--30.
https://doi.org/10.1016/j.ocecoaman.2012.12.002.

{[}36{]}{ }S. Gelcich, R. Guzman, C. Rodríguez-Sickert, J.C. Castilla,
J.C. Cárdenas, Exploring external validity of common pool resource
experiments: insights from artisanal benthic fisheries in Chile, Ecology
and Society. 18 (2013) art2. https://doi.org/10.5751/ES-05598-180302.

{[}37{]}{ }M. Andreu-Cazenave, M.D. Subida, M. Fernandez, Exploitation
rates of two benthic resources across management regimes in central
Chile: Evidence of illegal fishing in artisanal fisheries operating in
open access areas, PloS ONE. 12 (2017) e0180012.
https://doi.org/10.1371/journal.pone.0180012.

{[}38{]}{ }M. Scutari, C. Vitolo, A. Tucker, Learning Bayesian networks
from big data with greedy search: computational complexity and efficient
implementation, Stat Comput. 29 (2019) 1095--1108.
https://doi.org/10.1007/s11222-019-09857-1.

{[}39{]}{ }M. Scutari, Learning Bayesian Networks with the bnlearn R
Package, ArXiv:0908.3817 {[}Stat{]}. (2010).
http://arxiv.org/abs/0908.3817 (accessed December 10, 2020).

{[}40{]}{ }C.A. Chávez, J.J. Murphy, J.K. Stranlund, Managing and
defending the commons: Experimental evidence from TURFs in Chile,
Journal of Environmental Economics and Management. 91 (2018) 229--246.
https://doi.org/10.1016/j.jeem.2018.07.004.

{[}41{]}{ }SERNAPESCA, Fiscalización en Pesca y Acuicultura. Informe de
actividades del 2019, Ministerio de Economía Fomento y Turismo, Gobierno
de Chile, 2019.

{[}42{]}{ }X. Basurto, S. Gelcich, E. Ostrom, The social--ecological
system framework as a knowledge classificatory system for benthic
small-scale fisheries, Global Environmental Change. 23 (2013)
1366--1380. https://doi.org/10.1016/j.gloenvcha.2013.08.001.

{[}43{]}{ }L. Nahuelhual, G. Saavedra, G. Blanco, E. Wesselink, G.
Campos, X. Vergara, On super fishers and black capture: Images of
illegal fishing in artisanal fisheries of southern Chile, Marine Policy.
95 (2018) 36--45. https://doi.org/10.1016/j.marpol.2018.06.020.

{[}44{]}{ }AU-IBAR, Economic, Social and Environmental impact of
Illegal, Unreported and Unregulated Fishing in Africa, 2016.

{[}45{]}{ }A. Ospina-Alvarez, S. de Juan, K.J. Davis, C. González, M.
Fernández, S.A. Navarrete, Integration of biophysical connectivity in
the spatial optimization of coastal ecosystem services, Science of the
Total Environment. 733 (2020) 139367.
https://doi.org/10.1016/j.scitotenv.2020.139367.

{[}46{]}{ }S. de Juan, S. Gelcich, A. Ospina-Alvarez, A. Pérez-Matus, M.
Fernandez, Applying an ecosystem service approach to unravel links
between ecosystems and society in the coast of central Chile, Science of
The Total Environment. 533 (2015) 122--132.
https://doi.org/10.1016/j.scitotenv.2015.06.094.

{[}47{]}{ }K. Moon, D. Marsh, C. Cvitanovic, Coupling property rights
with responsibilities to improve conservation outcomes across land and
seascapes, CONSERVATION LETTERS. (2020).
https://doi.org/10.1111/conl.12767.

\appendix
\label{sec:appendix}


\section*{Supplementary material}

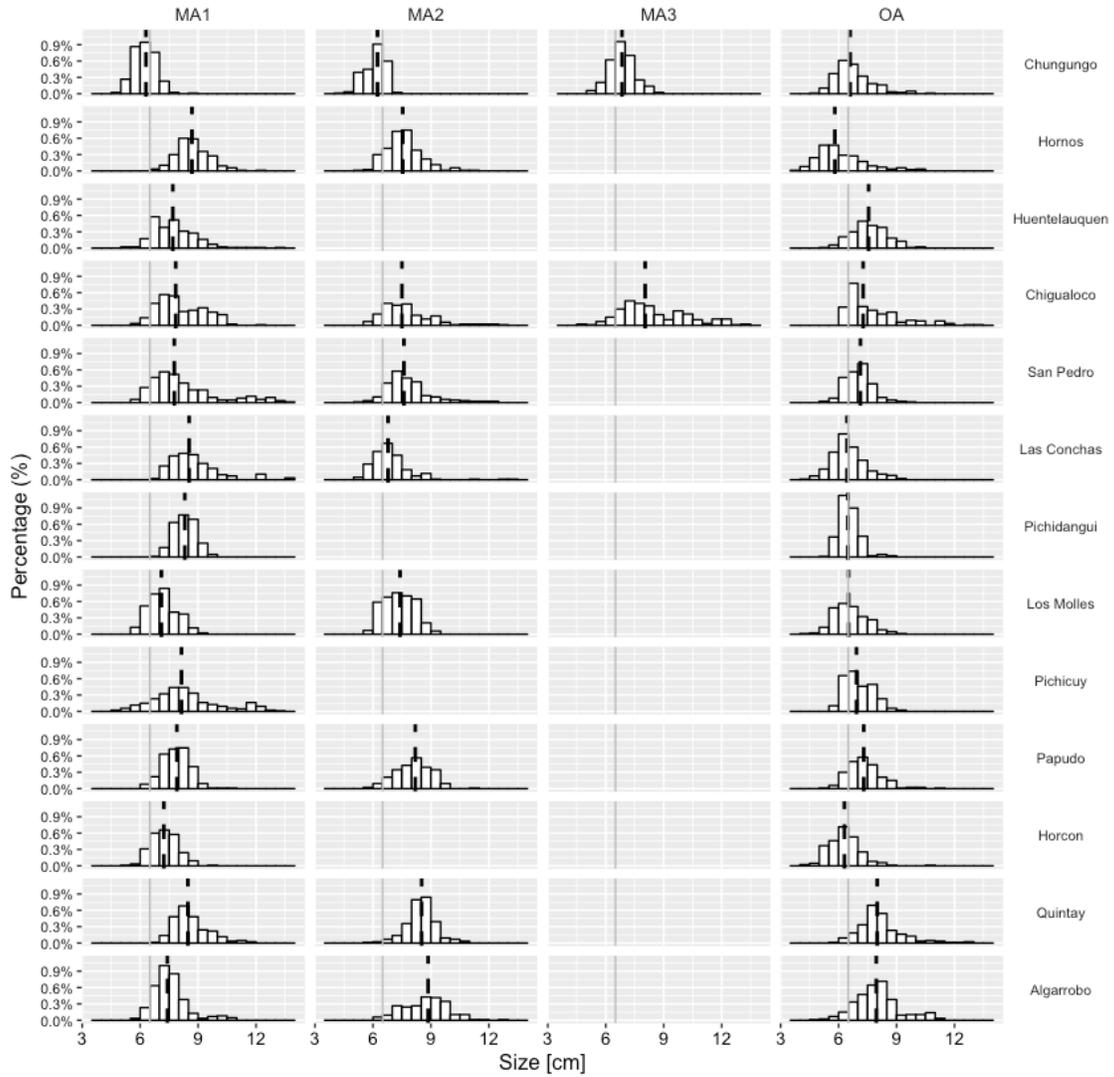

Figure S1 - Length-frequency distributions (in %) of keyhole limpets for the OA and different MAs and of each cove, sorted from north to south. Solid grey line indicates the MLS (=6.5 cm) for this benthic resource. Dashed black lines indicate the median size of the catches for each sampling site

Table S1 – Data gathered in the Management Areas and considered as potential drivers for illegal fishing in SSF in central Chile.

| Caleta | MA name | Nº MA per caleta | Access from land | Wave exposure | Surface MA (ha) | Other activities | Availability OA | Enforcement | Effectiveness | Distance from surveillance | Nº of poaching events |
|---|---|---|---|---|---|---|---|---|---|---|---|
| Chungungo | Chungungo E | 3 | Moderate | Protected | 21,45 | No | 0,02572 | low | low | far | <20 (low) |
| | Chungungo D | 3 | Difficult | Exposed | 18,82 | No | 0,02572 | low | very low | far | <20 (low) |
| | Temblador | 3 | Difficult | Moderate | 20,42 | No | 0,02572 | low | very low | far | <20 (low) |
| San Pedro – Los Vilos | Ñague | 2 | Moderate | Moderate | 280,77 | Yes | 0,03251 | very high | high | close | 50-100 (high) |
| | Ñague B | 2 | Difficult | Exposed | 56,49 | Yes | 0,03251 | very high | high | close | 50-100 (high) |
| Las Conchas | La Cachina | 2 | Moderate | Moderate | 116,2 | No | 0,03287 | high | moderate | close | 20-50 (moderate) |
| | Los Vilos Sector C | 2 | Easy | Moderate | 60,21 | No | 0,03287 | high | moderate | close | 20-50 (moderate) |
| Hornos | Hornos | 2 | Moderate | Moderate | 370,76 | No | 0,03496 | very high | moderate | average | >100 (very high) |
| | Hornos B | 2 | Difficult | Exposed | 84,32 | No | 0,03496 | very high | moderate | average | 50-100 (high) |
| Chigualoco | Caleta Boca del Barco | 3 | Moderate | Exposed | 112,17 | No | 0,03789 | moderate | low | close | 50-100 (high) |
| | Chepiquilla | 3 | Moderate | Exposed | 81,8 | No | 0,03789 | moderate | very low | close | >100 (very high) |
| | Chigualoco | 3 | Easy | Exposed | 343,11 | No | 0,03789 | moderate | low | close | >100 (very high) |
| Huentelauquén | Huentelauquén | 1 | Difficult | Exposed | 309,74 | No | 0,04815 | high | high | average | <20 (low) |
| Los Molles | Los Molles | 2 | Easy | Exposed | 95 | Yes | 0,10035 | very high | high | average | 20-50 (moderate) |
| | Playa Los Molles | 2 | Moderate | Exposed | 63,18 | Yes | 0,10035 | very high | high | average | 20-50 (moderate) |
| Algarrobo | Algarrobo A | 3 | Easy | Protected | 34,8 | No | 0,11637 | very high | moderate | close | 20-50 (moderate) |

| | | | | | | | | | | |
|---|---|---|---|---|---|---|---|---|---|---|
| | Algarrobo C | 3 | Easy | Exposed | 62,87 | No | 0,11637 | very high | high | close | 20-50 (moderate) |
| Pichidangui | Pichidangui | 2 | Easy | Exposed | 93,6 | Yes | 0,120033 | low | very low | close | >100 (very high) |
| Horcón | Horcón | 1 | Easy | Exposed | 98,71 | Yes | 0,13074 | low | very low | close | 50-100 (high) |
| Pichicuy | Pichicuy | 1 | Moderate | Exposed | 189,16 | Yes | 0,14445 | very high | moderate | close | 20-50 (moderate) |
| Papudo | Papudo | 2 | Moderate | Moderate | 187,26 | Yes | 0,1891 | very high | moderate | close | 50-100 (high) |
| | Punta Pite | 2 | Difficult | Exposed | 112,72 | Yes | 0,1891 | very high | low | close | 50-100 (high) |
| Quintay | Quintay A | 2 | Moderate | Protected | 53,42 | Yes | 0,34513 | very high | very high | average | >100 (very high) |
| | Quintay B | 2 | Moderate | Exposed | 105 | Yes | 0,34513 | very high | very high | average | >100 (very high) |

Table S2. Results from the Wilcoxon rank tests (W) comparing medians of the length of the catch of keyhole limpets (*Fissurella* spp.) between management regimes (MA indicates management areas and OA open access areas) at each cove (from south to north). N indicates sample size and includes the percentage of individuals below the minimum legal size (rounded). The index of availability of open areas is shown between parentheses for each cove. *P*-values are shown between parenthesis and values <0.05 in bold. Sites where the % of illegal catch (below MLS) was ≥ 49% are marked with a *.

| Cove | Site / Regime | Lapa (*Fissurella spp.*) | |
|---|---|---|---|
| | | N_total (N_undersize; % undersize) | W statistic for the catch (P value in parenthesis) |
| Algarrobo | MA 1- Sector A | 303 (20; 7%) | **31226 (9.105e-09)** |
| | MA 2- Sector C | 204 (5; 2%) | **39626 (4.027e-12)** |
| | OA | 284 (22; 8%) | - |
| Quintay | MA 1- Sector A | 215 (0; 0%) | **31031 (1.798e-08)** |
| | MA 2- Sector B | 245 (3; 1%) | **35041 (2.237e-08)** |
| | OA | 220 (9; 4%) | - |
| Horcón | MA 1- Horcón | 220 (27; 13%) | **40352 (< 2.2e-16)** |
| | OA * | 222 (131; 59%) | - |
| Papudo | MA 1– Papudo | 255 (2; 1%) | **35973 (5.339e-14)** |
| | MA 2– Pta. Pite | 217 (8; 4%) | **32034 (< 2.2e-16)** |
| | OA | 200 (28; 14%) | - |
| Pichicuy | MA 1– Pichicuy | 250 (28; 11%) | **45290 (< 2.2e-16)** |
| | OA | 240 (59; 25%) | - |
| Los Molles | MA 1– Los Molles | 272 (51; 19%) | **40408 (3.348e-14)** |
| | MA 2– Playa Los Molles | 322 (43; 13%) | **51168 (< 2.2e-16)** |
| | OA * | 212 (104; 49%) | - |
| Pichidangui | MA 1– Pichidangui | 224 (0; 0%) | **55114 (< 2.2e-16)** |
| | OA | 250 (0; 0%) | - |
| Las Conchas | MA 1– La Cachina | 241 (0; 0%) | **52041 (< 2.2e-16)** |
| | MA 2– Sector C | 207 (71; 34%) | **32832 (9.612e-06)** |
| | OA * | 256 (137; 53%) | - |
| San Pedro | MA 1- Ñague | 284 (27; 9%) | **41984 (< 2.2e-16)** |
| | MA 2– Ñague B | 204 (12; 6%) | **29778 (1.185e-13)** |
| | OA | 205 (41; 20%) | - |

| Site | Sample | Count (n; %) | Value (p) |
|---|---|---|---|
| Chigualoco | MA 1– Chigualoco | 255 (21; 8%) | **36099 (1.207e-07)** |
| | MA 2– Boca del Barco | 261 (13; 5%) | **35964 (2.935e-06)** |
| | MA 3– Chepiquilla | 181 (23; 13%) | 20818 (0.4811) |
| | OA | 221 (25; 11%) | - |
| Huentelauquen | MA 1– Huentelauquen | 225 (17; 8%) | 22904 (0.288) |
| | OA | 192 (23; 12%) | - |
| Hornos | MA 1– Hornos | 211 (0; 0%) | **39924 (< 2.2e-16)** |
| | MA 2– Hornos B | 263 (21; 8%) | **44646 (< 2.2e-16)** |
| | OA * | 201 (136; 68%) | - |
| Chungungo | MA 1– Chungungo E * | 271 (167; 62%) | **19143 (7.177e-10)** |
| | MA 2– Sector D * | 210 (149; 71%) | **13124 (7.108e-13)** |
| | MA 3– Temblador | 257 (66; 26%) | **30382 (0.01919)** |
| | OA | 210 (87; 41%) | - |

QUESTIONNAIRE: TURFs enforecement: implementation level and associated costs

**Date:**
**Cove:**
**Fisher's organization:**
**President's name:**
**TURF's name:**

1. Do you have hired a surveillance service?
   a. Yes
   b. No. Who does the surveillance?

2. Duration and frequency of surveillance
   a. Only during the day
   b. Day and night
   c. Only in weekdays
   d. Only with calm seas
   e. Other

3. This situation was different 10 years ago?

4. There is any protocol to carry out the surveillance? (e.g. patrol route, person who contact with NAVY)
   a. Yes. Who decided the protocol?
   b. No

5. Who pays the surveillance? Is the fisher's organization or another external entity?
   a. Fisher's organization
   b. External entity (Who?)
   c. Both

6. If the fisher's organizations must pay the surveillance. How much it cost?
   a. More than 75% of the organization budget
   b. More than 50% of the organization budget
   c. More than 25% of the organization budget
   d. Less than 10% of the organization budget

7. If the fisher's organizations have more than one TURF. The surveillance is the same for all?
   a. Yes
   b. No
      i. What TURF has more surveillance?
      ii. Why those differences?

8. In the last year. How many cases of poaching did the organization register in each of the TURFS?

9. In a scale 1 to 7. What is the effectivity of the surveillance?

10. In a scale 1 to 7. What is the importance of the surveillance in a TURF for its good working?

11. What need your organization to have a good surveillance?
    a. More money
    b. More cooperation between members
    c. Less necessity for people that need to poach
    d. SUBPESCA and SERNAPESCA provide mor staff in regions and coves
    e. Effective punishments



\end{document}